\documentclass[12pt,twoside]{article}
\usepackage{graphicx}
\usepackage[english]{babel}
\usepackage[cp866nav]{inputenc}

\newcommand{\bfi}{\begin{figure}}
\newcommand{\efi}{\end{figure}}

\newcommand{\non}{\nonumber \\}

\newcommand{\be}{\begin{equation}}
\newcommand{\ee}{\end{equation}}
\newcommand{\bea}{\begin{eqnarray}}
\newcommand{\eea}{\end{eqnarray}}
\newcommand{\lp}{\left (}
\newcommand{\rp}{\right )}
\newcommand{\lb}{\left \{}
\newcommand{\rb}{\right \}}
\newcommand{\lbr}{\left [}
\newcommand{\rbr}{\right ]}

\newcommand{\om}{\omega_0}

\newcommand{\ox}{x_0}

\newcommand{\ctnt}{\frac{C_t}{1+t}}
\newcommand{\csig}{\frac{1}{2\sigma\sqrt{\pi}}}
\newcommand{\inat}{\frac{i}{T}}
\newcommand{\zao}{\frac{1}{a_1}}
\newcommand{\zad}{\frac{1}{a_2}}
\newcommand{\sk}{\sigma^2}
\newcommand{\sch}{\sigma^4}
\newcommand{\sv}{\sigma^8}
\newcommand{\tk}{T^2}
\newcommand{\tc}{T^4}

\begin{document}


\title{Mutual space-frequency distribution of Gaussian signal}

\maketitle
\author{Yu. M. Kozlovskii}

\begin{abstract}
Mutual space-frequency distribution is proposed and it is shown
that Wigner and Weyl distribution functions are only particular
cases of these distribution. Mutual distribution for Gaussian signal
is analytically obtained. The simple connection between Wigner and
Weyl distributions is established. It is shown that Wigner
distribution forms as the rotational displacement of Weyl
distribution on informational diagram of conjugate coordinates
$(x;p)$ on an angle proportional to the mutual parameter $t$. The
results of direct calculations of mutual distribution for Gaussian
signal in the mutual domain are presented.
\end{abstract}




\section{Introduction}
The main aim of space-frequency analysis is elaboration of
distributions in order to get the information about signal
simultaneously in coordinate and frequency domains.As a rule the
Fourier transform is used to receive the signal frequency spectrum.
This well-known transformation is good tool for the analysis of
signal intensity distribution in the frequency plane. Such analysis
foresees the calculation of Fourier-spectrum for constant
coordinates. Practically, we have to deal with a certain momentary
value of coordinates for which the signal and their Fourier-image
are simultaneously determinated. With the coordinate variation the
appropriate conversion of Fourier-spectrum also takes place and at
once the problem of signal analysis occurs; the latter contains
frequency components that are variable in accordance with
coordinate.  In such a case it is important to know the value of
coordinate at which the corresponding transformation of frequency
spectrum takes place. In order to investigate the variation of
signal spectrum with the variation of its coordinate as far back as
in 60-80-ies of the previous century was propound a new approach. It
unites the information about coordinate and frequency constituent of
the signal in so-called space-frequency representations. In such
representations under consideration is a certain mutual function of
coordinate and frequency. The idea of construction of mutual
representations originates in works of E.Wigner (1932) \cite{wig},
D.Gabor (1946) \cite{gab} and J.Weyl (1932) \cite{weyl}. Before the
80-ies of the previous century tens of space-frequency
representations of such case were taken under consideration
\cite{coh,boa,coh2}. However, the Wigner and Weyl distributions that
are the most used for the present day remained the prerogative of
quantum mechanics and they have not have precisely expressed use.
Only in 1980 T.Klasen and B. Mecklenbrauker worked out the theory of
application of Wigner distribution for space-frequency analysis of
signals. Its main results were published in the series of works
under the title "Wigner distribution - the instrument for
space-frequency analysis of signals" \cite{cm1,cm2,cm3}. The
successful use of the Wigner distribution in the theory of signals
was stipulated by its  "good" mathematical characteristics,
especially by its representative characteristics that is basic in
the restoration of signal intensity distribution.

Within the framework of the given investigation among the variety of
space-frequency representations we single out two basic
distributions by Wigner and Weyl that are widely used by the theory
of signals in solving inverse physical problems \cite{boa,mer}. The
investigation of the signal characteristics takes place on the basis
of comparison with its displaced analogues. The shift within a time
results in subtraction of a specific value from the signal argument
\be \rightarrow x_{\tau}(t)=x(t+\tau). \label{zmx} \ee The suitable
displacement in accordance with frequency results in displacement of
argument of the Fourier-spectrum signal, what equals to
multiplication by phase multiplier in coordinate plane. \be {\bf
X_{\omega}} \rightarrow X_{\omega}(t)=x(t)e^{i\omega t}. \label{zmw}
\ee Similar correlations are well-known from classical analysis
\cite{mer}. Within the limits of space-frequency analysis we are
interested in the signals displaced simultaneously with time and
frequency, namely \be x_{-\frac{\tau}{2},-\frac{\omega}{2}}=x\lp
t-\frac{\tau}{2} \rp e^{-i\omega t/2}, \label{zmxwm} \ee \be
x_{\frac{\tau}{2},\frac{\omega}{2}}=x\lp t+\frac{\tau}{2} \rp
e^{i\omega t/2}. \label{zmxwp} \ee We can easily calculate the value
of displacement between the signals \be d({\bf
x_{-\frac{\tau}{2},-\frac{\omega}{2}}},{\bf
x_{\frac{\tau}{2},\frac{\omega}{2}}})^2= 2||x||^2-2\Re\lb {\cal
A}_{xx}(\omega,\tau) \rb. \label{disxw} \ee ${\cal
A}_{xx}(\omega,\tau)$ in this correlation plays a part of the
distance and is called a time-frequency autocorrelation function or
the ambiguity function
\begin{equation}
{\cal A}_{xx} (\omega,\tau) = \int\limits
x^*\left(t-\frac{\tau}{2}\right)
x\left(t+\frac{\tau}{2}\right)
\exp{(-i \omega t)}dt.
\label{weyl}
\end{equation}
In accordance to the Parceval theory we can rewrite this correlation by
Fourier-images of displaced signals
\begin{equation}
{\cal A}_{xx} (\omega,\tau) = \frac{1}{2\pi}\int\limits
X\left(\nu-\frac{\omega}{2}\right)
X^*\left(\nu+\frac{\omega}{2}\right) \exp{(i \omega \tau)}d\omega.
\label{weylf}
\end{equation}
The function in such a form was for the first time set by J. Weyl
\cite{weyl} and for the present time is known in the theory of
signals under the name of Weyl distribution or the ambiguity function.
Having realized the direct and inverse Fourier-transform we get the value
\begin{equation}
{\cal W}_{xx} (t,\nu) = \frac{1}{2\pi}
\int\limits_{-\infty}^{\infty}\int\limits_{-\infty}^{\infty} {\cal
A}_{xx} (\omega,\tau)e^{i\omega t}e^{-i\nu \tau}d\omega d\tau.
\label{weylwig}
\end{equation}
that in explicite form has the following notation
\begin{equation}
{\cal W}_{xx} (t,\nu) = \int\limits x^*\left(t-\frac{\tau}{2}\right)
x\left(t+\frac{\tau}{2}\right) \exp{(-i \nu \tau)}d\tau, \label{wig}
\end{equation}
or by the Fourier-image of the function  ${x}(t)$
\begin{equation}
{\cal W}_{xx} (t,\nu) = \frac{1}{2\pi}\int\limits
X\left(\nu-\frac{\omega}{2}\right)
X^*\left(\nu+\frac{\omega}{2}\right)
\exp{(i \omega t)}d\omega.
\label{wigf}
\end{equation}
The function was firstly introduced by E.Wigner and is named after
him - the Wigner function of distribution or just Wigner distribution \cite{wig}.

Joint space-frequency represenations are widely used not only in the
theory of signals. They have a number of practical use in different
fields of physics, geology, seismology, etc. Within the limits of the
given research we are interested in the use of such distributions in
the region of representations treatment and recognition of images.
This field of the physics imposes a set of demands that the generalized
distribution of signals have to meet. For the efficient use in the
theory of representatives of space-frequency distribution they have
to be characterized by representative property; for the limiting
values of the variable $t$ they have to develop into known distributions;
to have high distributive capacity in the field of Wigner distribution
as well as in Weyl distribution; to take positive values.

Representations that the most precisely meet the demands stated in
the theory of image are basic distributions by Wigner and Weyl.
These distributions also have their own peculiarities. Unfortunately,
up to the present there exists no simple deduction about the
expediency of use of this or that distribution. There exist a
number of signals for which Wigner and Weyl distributions proceed
into the negative region. In such cases these distributions are
interpreted as quasiprobable. In spite of the external resemblance
of the properties of Wigner and Weil distributions they have the
peculiarity of principle in the mechanism of renewal of the entrance
signal according to the known distribution. Wigner formalism allows
renewing the signal according to the so-called marginal distribution
\begin{equation}
|x(t)|^2=\int\limits_{-\infty}^{\infty}{\cal W}_{xx} (t,\nu)d\nu.
\label{marg}
\end{equation}
In the frame of Weyl formalism signal reconstruction takes place using
signal restoration scheme
\begin{equation}
|x(t)|^2=\int\limits_{-\infty}^{\infty}{\cal A}_{xx}
(0,\omega)e^{i\omega t}d\omega.
\label{schema}
\end{equation}
Sometimes working with the scheme of renewal according to the Weyl
distribution is to the great extent more easily (due to the
integrating of its crosscut) than in the case of marginal distribution.
For the present Wigner distribution is more commonly applied as it
uses marginal distributions that can be measured by experiment.
Though, both approaches have the right to existence.

Within the framework of the given experiment we investigate
uninterrupted transition between these distributions by means of
introduction of a generalized common function of time and frequency
that depends from a variable $t$.

For present day various types of space-frequency distributions are successfully
used for the analysis of nonstationary signals \cite{coh,boa}. Many of such
distributions are characterized by advantages as well as by disadvantages in
use in various fields of physics. Wigner and Weyl distributions are widely
used in space-frequency analysis and in particular in optical information
processing systems. Well known is a fact, that the very distributions posess
such charateristics, that are successfully used for description of many
optical systems. The spectrum of appliance of these distributions is
extremely wide. They are used, in particular, in the theory of optical lens
system, theory of communication, hydrolocation and other fields
\cite{cri,coh2,mer}. The researches of last year proved the efficiency of
use of space-frequency distributions in biology and medicine; especially
Wigner distribution was successfully used for renewal of volumetric structure
of objects within the framework of optical tomography \cite{kur,bre,smi}.
One of the promising investigation directions within the
space-frequency processing of signals is studying the properties of novel
space-frequency representations of the distributions, with the aim of their
further applications in different areas of physics and medicine.
Unfortunately, it often happens that some space-frequency distributions
do not meet demands raised by one or another specific application.
In this relation, many of the existing distributions need generalization
or improvements when applied to a given problem. During the second half
of the past century and the beginning of this one, a cleartendency has
been observed towards generalization of different space-frequency distributions.
The first attempt of such a generalization has been due to
L.Cohen \cite{coh3}  as long ago as in 1966. The author has introduced a
number of quasiprobable distributions that provide proper quantum mechanical
marginal distributions. Within the limits of this research the Wigner
distribution was examined as a separate case. The next step has been done
by N. De Brujin in 1973 \cite{de}. His work has been devoted to elaboration
of theory of generalized functions, with application concerned with Wigner
and Weil distributions. Summarizing the results of numerous investigations
L.Cohen \cite{coh,coh2} has suggested to has suggested a generalized
distribution involving a certain kernel
\bea
C(x,\omega,\Phi)&=&\frac{1}{2\pi}
\int\limits_{-\infty}^{\infty}\int
\limits_{-\infty}^{\infty}\int\limits_{-\infty}^{\infty}
f\lp y+\frac{x_0}{2} \rp f^*\lp y-\frac{x_0}{2} \rp \non &\times&
\Phi(x,\omega) e^{-i(\omega x_0-\om x +\om y)}dy dx_0 d\om.
\label{cohen}
\eea
Depending on the form of the kernel $\Phi(x,\omega)$ , this distribution
degenerates into one of the known distributions (Wigner, Weyl, Woodward,
Kirkwood, Page, Mark, etc.).

In the general case expression (\ref{cohen}) describes the class of
space-frequency distribution later named Cohen's class. Members of
this class are known distributions as well as a set of still unknown
distributions, that also satisfy all necessary requirements of existing
distributions. The theory of generalization of space-frequency distributions
has been developed by also famous specialist in the theory of signals by
A. Mertins. In his monograph "Signal Analysis"\cite{mer} he has singled
out thistopic into a separate section "General space-frequency distributions".
The author has stated that the Wigner distribution serves as an excellent
tool for space-frequency analysis as long as a linear dependence is
kept between the instantaneous coordinates andfrequencies. Otherwise,
a need in generalizing appears, whose general principles are
described in detail in the mentioned work.

Among numerous recent studies related to generalizing space-frequency
distributions, we should mention only the most typical ones.
The PhD Thesis by L. Durak "Novel time-frequency analysis technique
for deterministic signals" \cite{durak} is one of such studies, where
a close attention has been paid to generalizing distributions and
introducing their additional parameters.

Different types of generalizations of space-frequency distributions
have been thoroughly considered in the book by B. Boashash \cite{boa}.
Among a number of studies included in it, we would like to emphasize the
works by R. Baraniuk (p. 123), X.Xia (p. 223) and A. Papandreou-Suppappola
(p. 643). Within the mentioned collection, the work by G. Matz by F. Hlawatsch
(p. 400) is of particular interest in relation to the problem of
distributions generalization. It considers methodology for constructing
generalized distributions on the basis of both the Wigner distribution
and the ambiguity function (i.e. the Weyl distribution).

In the present work we try to use interlinks between the Wigner and
Weyl distributions with the aim of joining them into a single, more
general distribution. Up to date, it has been revealed that the two
distributions are related by a double Fourier transform. The results
obtained by us allow tracing transformation of one of
thedistributions into the other, while changing the distribution
parameter $t$. This generalized distribution generates a whole set
of new distributions formed in the process of switching between the
basic distributions. The latter fact may be important from the
viewpoint of possible practical applications. For the present day a
choice between the Weyl and Wigner distributions remains ambiguous.
Each of them has its own scheme for reconstruction of signal
intensity distribution. The scheme adopted for the Wigner
distribution includes calculating the marginal distributions
\cite{coh}. The Weyl distribution provides much simpler
reconstruction scheme,owing to simpler mathematical transformations
\cite{dopov,mv}. Traditionally, the Wigner distribution has been
used in a large majority of studies performed within the field.
Introduction of the mutual distribution would mean a possibility for
calculating 'mixed'states and determining the necessary
contributions of each of the limiting distributions. As stated
above, there appears a possibility for generalization of
distributions concerning various applied problems. However, only S.
Chountasis has suggested the approach \cite{cho} that enables
transitions between the Wigner and Weyl distributions. Such
distributions play an important role in the analysis of phase space
and, moreover, can be immediately applied in the Wigner tomography
\cite{kur,smi}..In 1999 S. Chountasis and co-authors \cite{cho,cho3}
have developed a general distribution based on the Wigner formalism,
which involves an additional parameter $\theta$. This study has been
performed in frame of quantum-mechanical formalism. It allows passing
the Wigner and Weyl distributions into each other by means of
changing the common parameter.

The problem of calculation of a classical analogue of this generalized
distribution remains urgent. It may be constructed based on the results
\cite{cho} or using the formalism of Weyl distribution, as has been done
by the present author when studying the properties of fractional Fourier
transform \cite{own}. Similarly to the works \cite{dopov,mv}, the author
has employed peculiarities of reconstruction of signal intensity based upon
the Weyl distribution. Meanwhile, it is just this reconstruction scheme is
realized experimentally in the real optical schemes \cite{own}.

\section{Mutual distribution: basic relations}

\subsection{Theoretical statements}
In the present work we propose to use a type of generalized
distribution with paramether $t$ based upon the Weyl
distribution function. The use of Weyl distribution has a peculiarity of
principle in comparison with the function of Wigner distribution,
that consists of the possibility to renew the intensity of
distribution. The latter is registered experimentally at the output
of the optical system. Common distribution of two signals $f_1(x)$
and $f_2(x)$ may be written as follows \cite{ufg}
\bea {\cal
K}_{f_1f_{2}^{*}}^{(t)}(x;p)&=& \frac{C_t}{1+t} \int\int d\ox d\om
\exp\lb i\lbr \ox p- \om x\rbr\rb \non &\times& \exp\lb
-i\frac{(x-\ox)^2+(p-\om)^2}{tan(\theta / \ 2)}\rb \non &\times&
\int\limits f_{1}\left(z+\frac{x_{0}}{2} \right)f_{2}^{*}
\left(z-\frac{x_{0}}{2}\right) \exp{(-i \omega_{0}z)}dz.
\label{gaf1}
\eea
Constant $C_t$ and variable of the generalized
distribution $t$ are determined by the expressions
\be
C_{t}=\frac{2}{\pi}\frac{1}{1-\exp{i\theta}},\qquad
t=\frac{\theta}{\pi}.
\label{const}
\ee
Distribution (\ref{gaf1}) is
called the mutual space-frequency distribution, or concisely the
mutual distribution. Limiting cases of ditribution (\ref{gaf1}) is
Weyl distribution (ambiguity function) (\ref{weyl}) with the
value of the parameter $t=0$ and Wigner distribution (\ref{wig})
with the value of parameter $t=1$. Accordingly, two known
distributions (\ref{weyl}) and (\ref{wig}) have plenty of
alternative distributions and to each of them corresponds a specific
value of parameter $t$.

The expression of the mutual distribution (\ref{gaf1}) may be also
represented in the simplified form using Weyl distribution
(\ref{weyl})
\bea
{\cal K}_{f_1f_2^*}^{(t)}(x;p)&=&\frac{C_{t}}{1+t}
\int\int d\ox d\om {\cal A}_{f_1f_2^*}(\ox;\om) \non &\times&
\exp\lb i\lbr \ox p- \om x\rbr\rb \exp\lb
-i\frac{(x-\ox)^2+(p-\om)^2}{tan(\theta /\ 2)}\rb.
\label{gaf}
\eea
Performing the converted transformation we can render the Weyl
distribution by means of the above introduced function of the mutual
distribution
\bea
{\cal A}_{f_1f_2^*}(\ox^{'};\om^{'})&=&\frac{1+t}{C_t} \int\int dx dp
{\cal K}_{f_1f_2^*}^{(t)}(x;p) \non &\times& \exp\lb -i\lbr \ox^{'}
p - \om^{'} x\rbr\rb \exp\lb
i\frac{(x-\ox^{'})^2+(p-\om^{'})^2}{tan(\theta / \ 2)}\rb.
\label{gafconj}
\eea
Formula (\ref{gafconj}) constitutes the
inverse connection between the simple ${\cal
A}_{f_1f_2^*}(\ox;\om)$ and generalized ${\cal
K}_{f_1f_2^*}^{(t)}(x;p)$ Weyl distributions. This makes the
possibility of restoring the distribution of signal intensity
according to the mutual distribution what has not been established
when using the generalized Wigner distribution \cite{cho}.

\subsection{Representation in the terms of Wigner distribution}
In order to make a comparison of the results with the existing
analogues it is indispensable to have possibility to precisely
calculate the limiting cases of the mutual distribution (17).
Calculation of the limiting case $t=1$ by means of the formula
(\ref{gaf}) may be conducted precisely, and in the case $t=0$ such
transition is not a trivia matter. To make the calculations simplier
we introduce the representation of the mutual space-frequency
distribution by means of Wigner distribution. We make use of the
known identity
\bea
{\cal A}_{f_1f_2^*}(\ox;\om)=\frac{1}{2\pi}\int\int d\xi d\eta
{\cal W}_{f_1f_2^*}(\eta;\xi)\exp(-i\om\eta)\exp(i\xi x_0),
\label{wv}
\eea
that connects Weyl and Wigner distributions.

Placing (\ref{wv}) into the expression (\ref{gaf}) and making a set
of conversions we receive a formula describing the mutual
space-frequency distribution in the terms of Wigner distribution
\bea
{\cal {K}}_{f_1f_2^*}^{(t)}(x;p)&=&\widetilde{C}_{t} \int\int d\ox d\om
{\cal W}_{f_1f_2^*}(\ox;\om) \exp\lb -i\lbr +\ox p+ \om x\rbr\rb \non
&\times& \exp\lb i\frac{1}{4}
tan\frac{\theta}{2}\lbr(x-\ox)^2+(p-\om)^2\rbr\rb,
\label{gaff}
\eea
where the constants $\widetilde{C}_t$ are determined by means of the
correlation
\be
\widetilde{C}_{t}=
\frac{-i}{\pi}\frac{1}{1-\exp{i\theta}}
\tan\frac{\theta}{2}\frac{1}{1+t}.
\label{const1}
\ee
As it can be easily seen in the case of
representation of the mutual distribution by means of the Wigner
distribution peculiarities in the point $t=0$ and around it
dissappear, however, there appear peculiarities around the point
$t=1$. Thereby the pair of representations: (\ref{gaf}) and
(\ref{gaff}) complement one another and completely describe mutual
space-frequency distribution in the region $t=[0,1]$.

\subsection{Limiting cases}
The aim of this work is to determine the mechanism of
re-distribution between Wigner and Weyl distributions and to
investigate the peculiarities of mutual distributions describing the
region of values $t=[0,1]$. Consequently the investigation of
limiting cases (\ref{gaf}) and (\ref{gaff}) of mutual distribution
is of peculiar importance. Let us study the limiting cases.

\emph{Case $t=1$.}

To describe this case we make use of coordinate representation of
the mutual distribution (\ref{gaf}). Placing $t=1$ (or $\theta=\pi$)
under (\ref{gaf}) we arrive to the following result
\bea
{\cal K}_{f_1f_2^*}^{t=1}(x;p)=\frac{1}{2\pi}\int\int d x_0 d\om {\cal
A}_{f_1f_2^*}(x_0;\om)\exp(i x_0 p-i\om x).
\label{gafw}
\eea
Accordingly to (8) we have
\bea
{\cal K}_{f_1f_2^*}^{t=1}(x;p)={\cal W}_{f_1f_2^*}(x;p).
\label{gafw1}
\eea
The limiting case $t=1$ of mutual space-frequency
distribution corresponds to the function of Wigner distribution.

\emph{Case $t=0$.}

To describe this case we make use of coordinate representation of
the mutual distribution (\ref{gaff}). Placing here $t=0$ (or
$\theta=0$) we arrive to the following result
\bea
{\cal K}_{f_1f_2^*}^{t=0}(x;p)=\frac{1}{2\pi}\int\int d x_0 d\om {\cal
W}_{f_1f_2^*}(x_0;\om)\exp(-i\om x+i x_0 p).
\label{gafv}
\eea
Accordingly to the formula connecting Wigner and Weyl distributions
we obtain
\bea
{\cal K}_{f_1f_2^*}^{t=0}(x;p)={\cal
A}_{f_1f_2^*}(x;p).
\label{gafv1}
\eea
Expression  (\ref{gafv1}) is
identical with the function of Weyl distribution. Hereby, introduced
by us distribution (\ref{gaf1}) in the values of the limiting cases
of the parameter $t$ describes known distributions (\ref{weyl}) and
(\ref{wig}). This distribution has two equivalent representations:
by means of the function of Wigner distribution (\ref{gaff}) or
function of Weyl distribution (\ref{gaf}). Having at least one from
the basic functions of distribution we can obtain the image of
mutual distribution. The object of the further investigation is to
study the peculiarities of the set of intermediate distributions
when $(0<t<1)$. In order to display the peculiarities of the mutual
distribution and to visually demonstrate the results we illustrate
it on the example of Gaussian signal. This is one of the simplest
types of signals that allows us to make calculations in the
explicit form. We shall notice that similar calculations may be
also made with other signals (in peculiar with orthogonal impulse,
etc.). Choice of the Gaussian signal is related to explicit form of
the common distribution. Herewith we can trace the mechanism of
transition of Weyl distribution into Wigner distribution and vice
versa.

\section{Mutual distribution of Gaussian signal}

\subsection{Basic relations}
In the work \cite{kozl} have been investigated the main
peculiarities of generalized coordinate-frequency distribution on
the example of Gaussian signal. Obtained results were based on the
frequent calculations of the proper distributions that allow to
properly evaluate processes of distribution. In the present work we
conduct an analytical calculation of the mutual space-frequency
distribution on the example of Gaussian signal that is represented
in the following way
\be
g(x)=\frac{1}{\sqrt{2\pi}\sigma}\exp{\lp-\frac{x^2}{2\sigma^2}\rp}.
\label{gf}
\ee
Fourier image of these function is
\be
\hat{{\cal F}}[g(x)]=\int\limits_{-\infty}^{\infty}g(x)e^{ixp}dx=\exp{\lp-\frac{p^2\sigma^2}{2}\rp}.
\label{fouriergf}
\ee
Such selection form of Gaussian functions are
stipulated by the fact that its integrating according to all values
$x$ results in value
\be
\int\limits_{-\infty}^{\infty}g(x)dx=1.
\label{normgf}
\ee
As investigations of the mutual space-frequency
distribution provides the study of re-distribution between Wigner
and Weyl distributions it is reasonable to present the explicit form
of these distributions for the case of Gaussian signal \cite{kozl}.

\emph{Weyl distribution for Gaussian signal }

\be
{\cal A}_{ff^{*}}(x_{0};\omega_{0})=\frac{1}{2\sqrt{\pi}\sigma}
\exp{\lp-\frac{x_0^2}{4\sigma^2}-\frac{\om^2\sigma^2}{4}\rp}.
\label{weylgf}
\ee

\emph{Wigner distribution for Gaussian signal}

\be
{\cal W}_{ff^{*}}(x;\omega)=\frac{1}{\sqrt{\pi}\sigma}
\exp{\lp-\frac{x^2}{\sigma^2}-\omega^2\sigma^2\rp}.
\label{wignergf}
\ee

It is known \cite{coh}, that one of the basic characteristics of the signal
is its representativity that is the possibility to restoring the signal according
to its distribution. Schemes of renewing for Weyl and Wigner distributions
are well known. For the case of Gaussian signal they have the following form

\emph{restoration scheme for Gaussian signal by Weyl distribution}

\be
|g(x)|^2=\frac{1}{2\pi}\int\limits_{-\infty}^{\infty}{\cal
A}_{ff^{*}}(0;\omega_{0})e^{i\om x}d\om=
\frac{1}{2\pi\sigma^2}\exp{\lp-\frac{x^2}{\sigma^2}\rp}.
\label{rsweylgf}
\ee

\emph{restoration scheme for Gaussian signal by Wigner distribution}

\be
|g(x)|^2=\frac{1}{2\pi}\int\limits_{-\infty}^{\infty}{\cal
W}_{ff^{*}}(x;\omega)d\omega=
\frac{1}{2\pi\sigma^2}\exp{\lp-\frac{x^2}{\sigma^2}\rp}.
\label{rswignergf}
\ee
Independently from the choice of distribution form Gaussian signal is precisely
renewed using both schemes of renewing.

We find explicit analytical form of the mutual distribution for the
case of Gaussian signal. For this reason we will place into the
formula of mutual distribution (\ref{gaf}) the Weyl distribution of
Gaussian signal (\ref{gaf}). We obtain expression
\bea
{\cal K}_{ff^*}^{t}(x;p)=\ctnt\csig\exp\lb -\inat(x^2+p^2) \rb
I_1(x,p)I_2(x,p),
\label{mgdgen}
\eea
where
\bea
I_1(x,p)=\int
dx_0\exp\lb i x_0 p+\inat 2xx_0-\inat x_0^2-\frac{x_0^2}{4\sigma^2}
\rb,
\label{i1}
\eea
\bea
I_2(x,p)=\int d\om\exp\lb -i\om x+\inat
2p\om-\inat \om^2-\frac{\om^2\sigma^2}{4} \rb.
\label{i2}
\eea
Expressions (\ref{i1}) and (\ref{i2}) may be written in the
following form
\bea I_1(x,p)=\frac{\sqrt{\pi}}{\sqrt{a_1}} \exp\lb
-\zao \lp \frac{x}{T}+\frac{p}{2} \rp^2 \rb,
\label{i1n}
\eea
\bea
I_2(x,p)=\frac{\sqrt{\pi}}{\sqrt{a_2}} \exp\lb -\zad \lp
\frac{p}{T}-\frac{x}{2} \rp^2 \rb,
\label{i2n}
\eea
where the following symboles are introduced
\bea
a_1=\frac{T+i4\sigma^2}{4T\sigma^2}, \qquad
a_2=\frac{T\sigma^2+4i}{4T}, \qquad T=\tan\lp\frac{\theta}{2}\rp.
\label{pozng}
\eea
The mutual distribution of Gaussian signal (\ref{mgdgen}) has the form
\bea
{\cal K}_{ff^*}^{t}(x;p)&=&
C_{t}^{\sigma} \exp\lb -\inat\lp x^2+p^2 \rp \rb \non &\times&
\exp\lb -\zao\lp \frac{x}{T}+\frac{p}{2} \rp^2 \rb \exp\lb -\zad\lp
\frac{p}{T}-\frac{x}{2} \rp^2 \rb.
\label{mgd}
\eea
where constant
\be
C_{t}^{\sigma} =
\ctnt\frac{\sqrt\pi}{2\sigma}\frac{1}{\sqrt{a_1a_2}}
\label{cmgd}
\ee
depends from dispersion of Gaussain distribution of signal
$\sigma$ and from the values $a_1$, $a_2$. From the values
$a_1^{-1}$ and $a_2^{-1}$ depend also expressions in the index of
the expression exponential curve (41). We depict them in the form
\be
a_{1}^{-1}=a_{11}+ia_{12}:  \qquad
a_{11}=\frac{4\tk\sk}{16\sch+\tk},\qquad
a_{12}=-\frac{16T\sch}{16\sch+\tk},
\label{a1c}
\ee
\be
a_{2}^{-1}=a_{22}+ia_{21}:  \qquad
a_{22}=\frac{4\tk\sk}{16+\tk\sch},\qquad
a_{21}=-\frac{16T}{16+\tk\sch}.
\label{a2c}
\ee
Placing (\ref{a1c})
and (\ref{a2c}) under (\ref{mgd}) we find explicit analytical form
of the mutual distribution of the Gaussian signal (\ref{gaf1})
\bea
{\cal
K}_{ff^*}^{t}(x;p)&=&\ctnt\frac{\sqrt(\pi)}{2\sigma}\frac{1}{\sqrt{a_1a_2}}
\exp\lb -\inat\lp x^2+p^2 \rp \rb \non &\times& \exp\lb -a_{11}\lp
\frac{x}{T}+\frac{p}{2} \rp^2-ia_{12}\lp \frac{x}{T}+\frac{p}{2}
\rp^2 \rb \non &\times& \exp\lb -a_{22}\lp \frac{p}{T}-\frac{x}{2}
\rp^2 -ia_{21}\lp \frac{p}{T}-\frac{x}{2} \rp^2 \rb.
\label{mgdlc}
\eea
We may check that expression (\ref{mgdlc}) in limiting cases
$t=0$ and $t=1$ changes into Weyl and Wigner distributions
respectively. In order to ascertain the circumstance we investigate
the conduct of the expression (\ref{mgdlc}) in the limit
$t\rightarrow 1$. Parameter $t$ is determined according to
(\ref{const}) by value $\theta$. We introduce small value $\alpha$
($\alpha\ll 1$) and depict $\theta$ in the form \be \theta = \pi -
\alpha. \label{tetam} \ee Region of small values $\alpha$
corresponds to quasi-Wigner region of mutual function of
distribution. Then the value $C_t$ that is a part of
$C_{t}^{\sigma}$ in case of small values $\alpha$ can be depicted as
follows
\be
C_t \approx \frac{1}{\pi} \left( 1 + i \frac{\alpha}{2}
\right).
\label{ct}
\ee
Asymptotics of the values $(a_1a_2)^{-1/2}$
may be easily find when use its module $r_{12}$ and argument
$\varphi_{12}=\varphi_1+\varphi_2$
\be
(a_{1}a_{2})^{-1/2}=r_{12} e^{i(\varphi_1+\varphi_2)/2},
\label{a12}
\ee
where
\bea
r_{12}=\pm\frac{4
T\sigma}{(16+\tk\sch)^{1/4}(16\sch+\tk)^{1/4}},\non
\varphi_{1}=\arctan \lp-\frac{4\sk}{T}\rp, \quad \varphi_{2}=\arctan
\lp-\frac{4}{T\sk}\rp.
\label{r12}
\eea
In the case of small values
$\alpha$ ($T\gg 1$) we have such approximate form for the value
$r_{12}$
\be
r_{12} = 4 \lp 1 - \frac{4}{T^2} \lp \sigma^4 +
\sigma^{-4}\rp\rp,
\label{r12m}
\ee
and for value $\varphi_1$ and $\varphi_2$ we find
\be
\varphi_{12} = \varphi_1 + \varphi_2 = -
\frac{4}{T} \lp \sigma^2 + \sigma^{-2} \rp.
\label{fi12m}
\ee
Thereby, the constant $C_{t}^{\sigma}$ from (\ref{mgdlc}) in
quasi-Wigner region $\alpha \ll 1$ has the form
\be
C_{t}^{\sigma(1)} = \frac{1}{\sigma\sqrt\pi} \lp 1 + \frac{i}{2}
\alpha \rp \lp 1 - \frac{4}{T^2} \lp \sigma^4 + \sigma^{-4}\rp\rp
e^{-\frac{2i}{T} \lp \sigma^2 + \sigma^{-2}\rp}.
\label{ctmal}
\ee
Index 1 in the value $C_{t}^{\sigma}$ denotes a condition $\alpha\ll
1$. Naturally, that in the $\alpha\rightarrow 0$ we has the value
\[C_{t}^{\sigma} = \frac{1}{\sigma \sqrt\pi},\] that precisely corresponds
to the amplitude of the expression for Wigner distribution (\ref{wignergf}).

Let us observe the coefficient $C_{t}^{\sigma}$ from (\ref{mgdlc})
in the case of small values $\theta$ $(t\rightarrow 0)$. We shall
call the region of parameter $t$ values quasi-Weyl region as far as
in the value $t=0$ we aquire Weyl distribution. Similarly as in the
case $t=1$, we find for small values the following expressions
\be
C_{t}=\frac{i}{\pi T}(1-i T), \qquad r_{12}=T, \qquad
\varphi_1=\varphi_2=-\pi/2, \qquad
C_{t}^{\sigma(2)}=\frac{1}{2\sqrt{\pi}\sigma}.
\label{ctst0}
\ee
Such value of constant $C_{t}^{\sigma(2)}$ exactly corresponds to
the amplitude of the value from Weyl distribution of Gaussian signal
(\ref{weylgf}). Thus, the mutual distribution of Gaussian signal in
limiting cases according to amplitude precisely coincides with the
known distributions of Wigner and Weyl.

Calculation of real and imaginary part of constant $C_{t}^{\sigma}$
depicted on the Fig.1. for $C_{t}^{\sigma}$ provides an exact
correspondence according to amplitude with Wigner and Weyl
distributions. In the common region appears imaginary part
$C_{t}^{\sigma}$ that is inherent only in mutual distribution. In
the limiting cases $t=0(\theta=0^{0})$ and $t=1(\theta=180^{0})$
imaginary part dissappears what corresponds to the cases of basic
distributions. Worth mentioning is also peculiar conduct of the
imaginary part of constant $C_t^\sigma$ having maximum in the
quasi-Wigner region.

Let us proceed the investigation of limiting cases of expessions
placed in the index of the exponent on a curve formula
(\ref{mgdlc}). As was shown above the amplitude of mutual
distribution in limiting cases coincides with the amplitude of known
distributions of Weyl and Wigner. Providing that the form of these
dirtibutions will also coincide functionally (as functions $x$ and
$p$), mutual distribution may be considered as generalization of
well-known distributions of Wigner and Weyl. Having made a number of
mathematical transformations the expression (\ref{mgdlc}) aquires
the following form
\bea
{\cal K}_{rr^*}^{(t)}(x;p)&=& C_{t}^{\sigma}
\exp\lb - g x^2 - f p^2 - d x p \rb,
\label{Kgfd}
\eea
where
coefficients $g$, $p$ and $d$ are certain complex values
\be
g = g_0+i g_1, \quad f = f_0+i f_1, \quad d = d_0+i d_1
\label{gfd}
\ee
moreover
\bea
&&g_0=\frac{\sk}{\mu}(64+20\tk\sch+\tc), \qquad
g_1=\frac{T}{\mu}(16[1-4\sch]+\tk[\sch-4]);\non
&&f_0=\frac{\sk}{\mu}(64\sch+20\tk+\tc\sch), \qquad
f_1=\frac{T\sch}{\mu}(16[\sch-4]+T[1-4\sch]);\non
&&d_0=\frac{4T\sk}{\mu}(1-\sch)(16-\tk), \qquad
d_1=16\tk\frac{1-\sv}{\mu}.
\label{pozn}
\eea
where the value $\mu$ has the form
\bea
\mu=16^2\sch +16\tk(1+\sv)+\sch\tc
\label{mu}
\eea
Constant $C_{t}^{\sigma}$ from (\ref{cmgd}) may be represented as
\bea
C_{t}^{\sigma}=
\frac{i}{\pi}\frac{1+e^{-i\theta}}{\sin\theta}\frac{1}{1+t}
\frac{\sqrt{\pi}}{2\sigma}r_{12}e^{i(\varphi_1+\varphi_2)/2},
\eea
where
\bea r_{12}=\frac{4\sigma
T}{(16+\tk\sch)^{1/4}(16\sch+\tk)^{1/4}}, \non \varphi_1=\arctan\lp
-\frac{4\sk}{T} \rp, \qquad \varphi_2=\arctan\lp -\frac{4}{T\sk} \rp
\eea
Representation (\ref{Kgfd}) is an explicit form of the mutual
distribution of Gaussian signal. Let us consider asymptotic values
of $g$, $f$ and $d$ in the limiting cases $t=0$ and $t=1$.

In the case $t=0$ ($T\rightarrow 0$) from (\ref{pozn}), (\ref{mu}) we find
\be
g_0 = (4\sigma^2)^{-1}, \quad f_0 = \sigma^2/4.
\ee
The rest of coefficients turns into zero
\be
g_1 = f_1 = d_0 = d_1 = 0.
\label{g1f1}
\ee
In the limiting case $t=1$ ($T\rightarrow \infty$) we receive
\be
g_0 = \sigma^{-2}, \quad f_0 = \sigma^2.
\ee
Other coefficients dissapear as in (\ref{g1f1}). We shall notice that imaginary
part of cross-expressions in (\ref{Kgfd}) real part dissapears in $t\rightarrow 0$
as well as in $t\rightarrow 1$.

What concerns the values $g$ and $f$, their imaginary parts also tend
to zero in the case $t=0$ and $t=1$. Thus, the very quadratical
members are responsible for forming of the distribution in limiting
cases as they are included into known Wigner and Weyl distributions,
and cross-representations that are inherent only in mutual
distribution dissapear.

\section{Signal restoration scheme in mutual distribution domain}

As it is known, restoration of the signal according to Weyl
distribution takes place correspondingly to (\ref{rsweylgf}), and
according to Wigner distribution - accordning to (\ref{rswignergf}).
Above mentioned schemes of signal restoration may be united into one
formula using the mutual space-frequency distribution suggested
above. We shall introduce the value
\be
f_\theta (x) =
\frac{1}{2\pi} \int_{-\infty}^{\infty} d \rho w_\theta \lp x \sin
(\theta/2) p \rp e^{ipx\cos(\theta/2)}.
\label{fteta}
\ee
It can be easily seen, that when $\theta=0$ we arrive at Weyl renewal scheme,
and when $\theta=\pi$ - we aquire Wigner renewal signal. Taking into
consideration (\ref{Kgfd}), for the function $w_\theta$ we have
\bea
w_\theta \lp x \sin (\theta/2),p \rp = C_{t}^{\sigma} \exp\lb - g
\sin^2 (\theta/2) x^2 - f p^2 - d \sin (\theta/2) x p \rb.
\label{ftetai}
\eea
In the result of integration of (\ref{fteta}) we aquire
\bea
f_\theta(x) &=& \frac{C_{t}^{\sigma}}{2\pi}
\frac{\sqrt\pi}{(f_0+if_1)^{1/2}} e^{-x^2\sin \theta/2 (g_0+ig_1)}\non &\times&
\exp \lb \frac{x^2}{4} \frac{1}{(f_0+if_1)} \lp \sin
\frac{\theta}{2} (d_0+id_1) - i\cos \theta/2 \rp^2 \rb.
\label{ftetaya}
\eea
It can be easily seen that in limiting cases
function $f_\theta(x)$ transforms into expressions (\ref{rsweylgf})
and (\ref{rswignergf}). Taking into account expressions for real and
imaginary part of coefficients $g$, $f$ and $d$ we arrive at
\be
f_\theta (x) = \frac{1}{2\sqrt{\pi r_f}} e^{-\frac{i}{2}\varphi_f}
e^{-Gx^2}, \label{rs}
\ee
where
\[
r_f = \frac{\sigma^2}{\mu} \lbr \lb 64\sigma^4 + 20 T^2 + \sigma^4 T^4 \rb
+ T^2\sigma^4 \lb 16 (\sigma^4-4) +T^2 (1-4\sigma^4)\rb^2 \rbr^{1/2},
\]
\[
\varphi_f = \arctan \lp \sigma^2 T \frac{16(\sigma^4-4)+T^2(1-4\sigma^4)}{64\sigma^4+20T^2+T^4\sigma^4}
\rp,
\]
\[
G = r_g \sin \frac{\theta}{2} e^{i\varphi_g} - \frac{1}{4} \frac{1}{r_f}
\lb r_\alpha^2 \sin^2 \frac{\theta}{2} e^{i(\varphi_f+2\varphi_d)} -
2 r_d \sin \frac{\theta}{2} e^{i(\varphi_f+\varphi_d+\frac{\pi}{2})} -
\cos^2 \frac{\theta}{2} e^{i\varphi_f}\rb.
\]
In order to make the formula shorten certain symbols are introduced
\[
r_g = \frac{\sigma^2}{t_4} \lbr \lb 64 + 20T^2 \sigma^4 +T^4 \rb + \frac{T^2}{\sigma^4}
\lb 16 (1-4\sigma^4) + T^2 (\sigma^4-4)\rb^2\rbr^{1/2},
\]
\[
\varphi_g = \arctan \frac{g_1}{g_0} = \arctan \lb \frac{T[16(1-4\sigma^4)+T^2(\sigma^4-4)]}
{\sigma^2(64+20T^2\sigma^4+T^4)} \rb,
\]
\[
r_d = \frac{4T}{t_4} \lbr \sigma^4 (16-T^2) (1-\sigma^4)^2 + 16 T^2 (1-\sigma^8)^2 \rbr^{1/2},
\]
\[
\varphi_d = \arctan \lp \frac{16T^2(1-\sigma^8)}{4T\sigma^2(1-\sigma^4)(16-T^2)} \rp =
\arctan \lp 4T \frac{(1-\sigma^8)}{\sigma^2(1-\sigma^4)(16-T^2)}\rp.
\]
Thus, suggested restoration scheme (\ref{rs}) in limiting cases precisely renew
Gaussian signal (\ref{gf})
\be
f_{(t=1)} (x)= f_{(t=0)} (x) = \frac{1}{\sqrt{2\pi}\sigma}\exp{\lp-\frac{x^2}{2\sigma^2}\rp}.
\label{rslimits}
\ee
Thereby, side by side with known limiting cases arises possibility of
signal intensity distribution restoration in the region $t=[0..1]$. In our opinion investigation
of intensity distribution in the mentioned region is an urgent problem, however
it passes the limits of the present investigation.

\section{Conclusions}

In this work we propose a mutual space-frequency distribution as
generalization of Weyl distribution (\ref{gaf}). The mutual distributions is
characterized by a certain parameter $t$. It is a generalization of
space-frequency representations suggested by Wigner and Weyl and
comprises them as limiting cases. In the course of the investigation
it has been determined that transition from Weyl distribution into
Wigner one occurs by means of mutual region as a curve at the
information diagram of mutual coordinates $(x,p)$ Fig.2. When
modificating the mutual parameter $t$ the distribution changes and
simultaneously suffers deformation for the angle propotional to
parameter $t$. Fig.3 depicts the region of mutual distribution close
to Weyl distribution (quasi-Weyl region of distribution). Fig. 3(a)
illustrates Weyl distribution of Gaussian signal that is formed from
the mutual distribution (\ref{Kgfd}) when the value of mutual
parameter $t=0$. Increase of value of mutual parameter till $t=0,1$
leads to the curve of the mutual distribution at informational
diagram (Fig.3(b)). When $t=0,25$ beside curve also starts the
pocess of deformation that leads to the transformation of Weyl
distribution into Wigner distribution (Fig.3(c)). The peculiar is
the value of parameter $t=0,5$. mutual distribution in this case is
placed precisely in the middle between limiting cases of Weyl and
Wigner distributions (Fig.5). In the process of increase of the
mutual parameter $t$ a transformation of Weyl distribution into
Wigner distribution takes place by means of change of the curve
counterclockwise to the mutual space-frequency distribution. In the
limiting case $t=1$ from the mutual distribution Wigner distribution
is formed (Fig.4(a)). When the mutual parameter $t$ decreases in the
region of Wigner distribution the curve at the informational diagram
of joined coordinates $(x,p)$ is changed (Fig.4(b,c)). In the region
of Wigner distribution when the parameter $t$ decreases the curve of
mutual distribution is changed clockwise. When the value $t=0,5$ the
mutual distribution is formed what can be observed at Fig.5.
Thereby, we come to conclusion that in the process of changing the
parameter $t$ of the mutual distribution the Weyl and Wigner
distributions move towards one another at informational diagram and
are put in equilibrium in the point $t=0,5$ (Fig.5). It can be
stated that Wigner distribution is formed as a change of curve of
Weyl distribution at informational diagram for the angle
$\theta<90^{0}$. Similarly, Weyl distribution is formed as a change
of curve of Wigner distribution in the contrary way. It should be
noticed that in the mutual region the distribution becomes complex
one (Fig.6(b-d)), (Fig.7(b-d)). However, in the known liniting cases
only the real part has a contribution: $t=0$ (Fig.6(a)),(Fig.7(a))
and $t=1$ (Fig.6(e)),(Fig.7(e)).

In this paper we propose new space-frequency distribution (\ref{gaf}), which
could play an important role in the optical information processing schemes describing.

\newpage

\begin{figure}[htbp]
\includegraphics[width=1.0\textwidth]{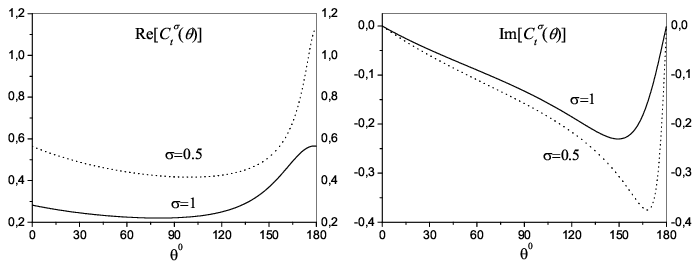}
\caption{Real and Image parts of the Gaussian mutual distribution
constant $C_{t}^{\sigma}$.}
\label{fig1}
\end{figure}

\newpage

\begin{figure}[htbp]
\includegraphics[width=0.55\textwidth]{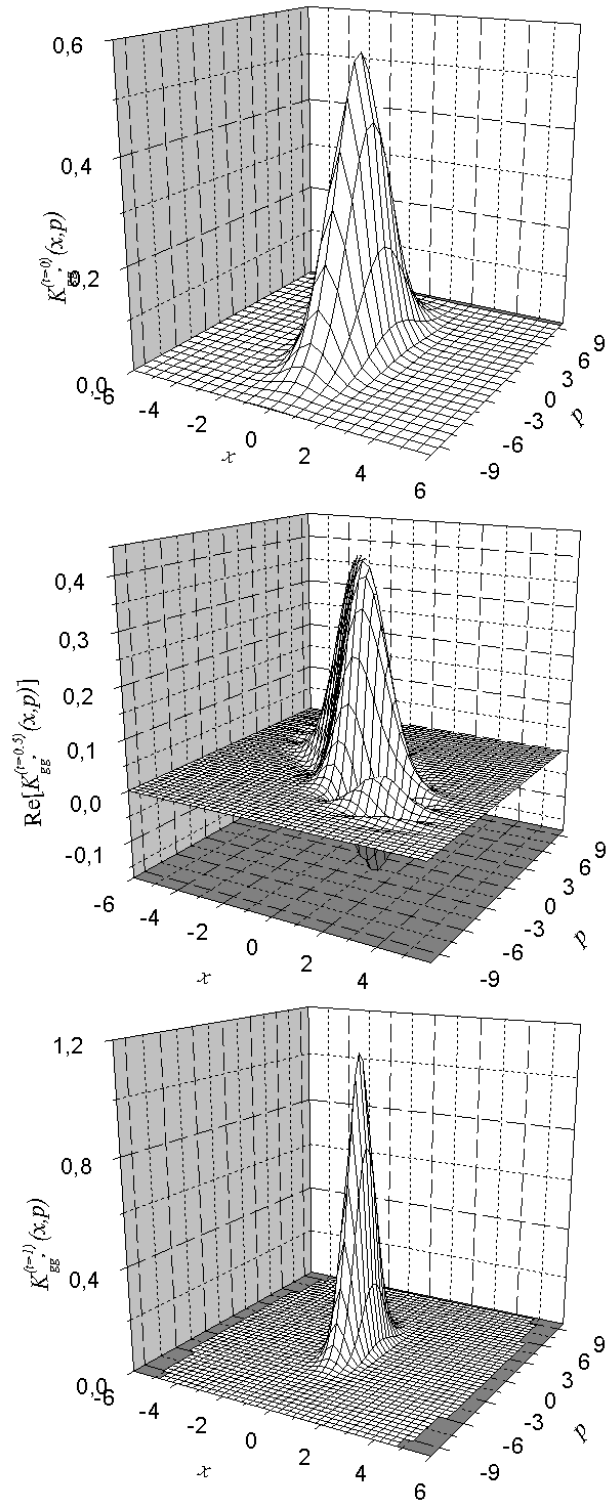}
\caption{Redistribution of the Gaussian mutual distribution from
Weyl to Wigner domain.}
\label{fig2}
\end{figure}

\newpage

\begin{figure}[htbp]
\includegraphics[width=0.4953\textwidth]{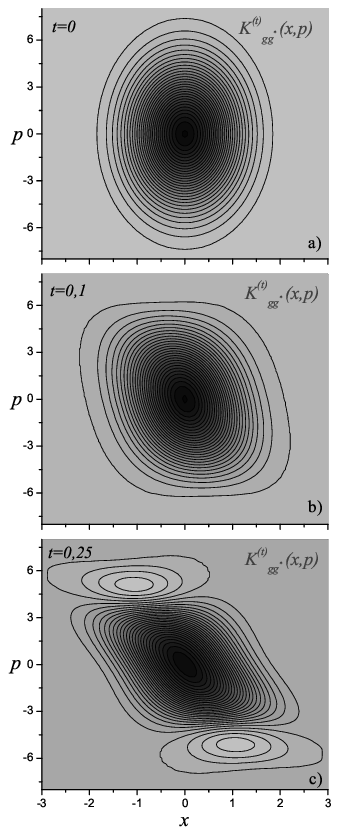}
\caption{Rotational displaysment of the Gaussian mutual distribution
in the Weyl domain at different values of mutual parameter $t$ (Real
part).} \label{fig3}
\end{figure}

\newpage

\begin{figure}[htbp]
\includegraphics[width=0.4953\textwidth]{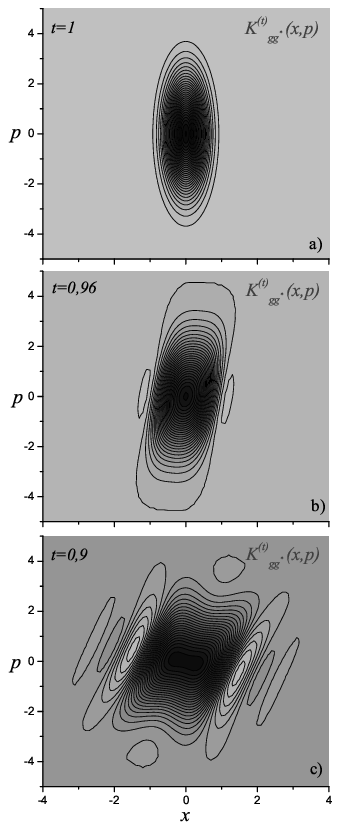}
\caption{Rotational displaysment of the Gaussian mutual distribution
in the Wigner domain at different values of mutual parameter $t$
(Real part).} \label{fig4}
\end{figure}

\newpage

\begin{figure}[htbp]
\includegraphics[width=0.9\textwidth]{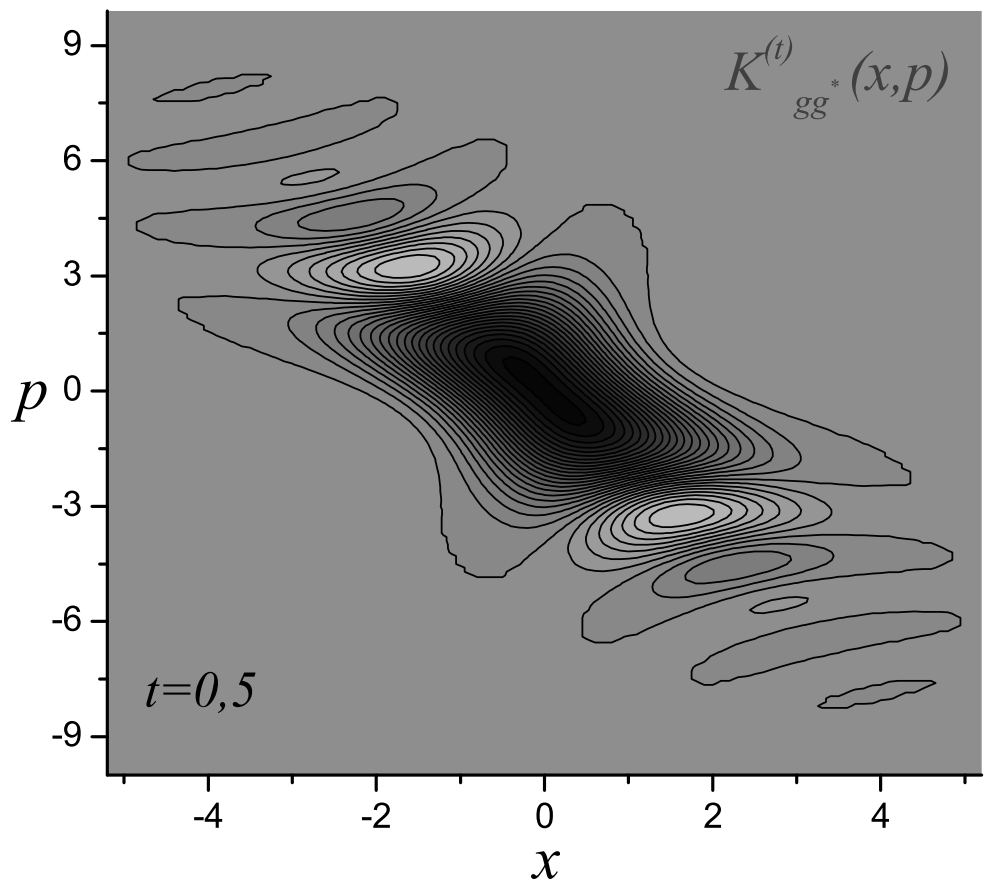}
\caption{Mutual distribution of the Gaussian signal
at the value of mutual parameter $t=0.5$ (Real part).}
\label{fig5}
\end{figure}

\newpage

\begin{figure}[htbp]
\includegraphics[width=0.35\textwidth]{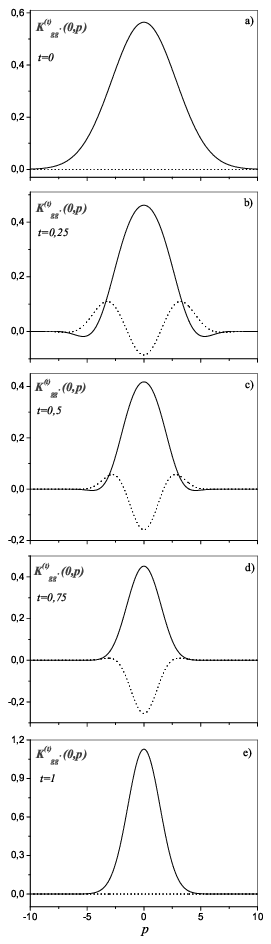}
\caption{Mutual space-frequency distribution in direction $x=0$ of Gaussian signal
at different values of mutual parameter $t$, solid line real part and dot line image part.}
\label{fig6}
\end{figure}

\newpage

\begin{figure}[htbp]
\includegraphics[width=0.35\textwidth]{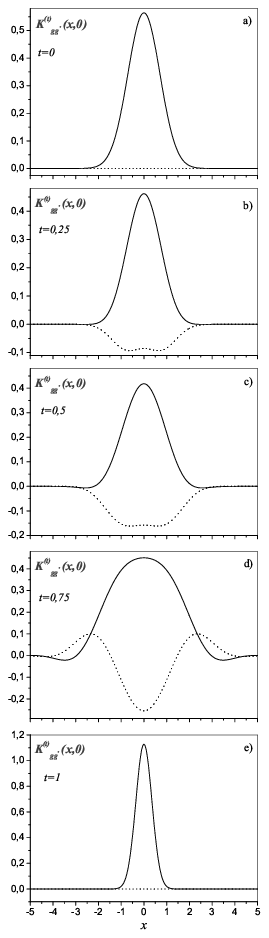}
\caption{Mutual space-frequency distribution in direction $p=0$ of Gaussian signal
at different values of mutual parameter $t$, solid line real part and dot line image part.}
\label{fig7}
\end{figure}

\end{document}